\documentstyle[12pt,epsfig]{article}
\begin{document}
\baselineskip=20pt

\title{$D \to PV$ decays with Final State Interactions
   \thanks{Supported in part by National Natural Science Foundation of
   China.}
   \thanks{ lijw@mail.ihep.ac.cn (J.W. Li); yangmz@mail.ihep.ac.cn (M.Z. Yang);
     duds@mail.ihep.ac.cn (D.S. Du).}}
\author{Jing-Wu Li$^b$, Mao-Zhi Yang$^{a,b}$, Dong-Sheng
Du$^{a,b}$\\[4mm]
       $^a$ CCAST(World Laboratory), P.O.Box 8730, Beijing 100080, China\\
$^b$ Institute of High Energy Physics, Chinese Academy of
             Sciences, P.O.Box 918(4),\\ Beijing 100039, China }

\maketitle
\begin{picture}(0,0)
\put(355,255){BIHEP-TH-2002-23}
\end{picture}

\vspace{2cm}

\begin{abstract}
We employ one-particle-exchange method to study $D  \to PV $
decays in $D\to K \rho$, $\pi K^*$, $\pi\rho $ processes.   Taking into account a strong
phase  and considering nonfactorizable effect,  we can get good results
consistent with the experimental data.  Nonfactorizable effect is not always
large, but in some cases, the nonfactorizable effect is necessary to accommodate
the  experimental data. Strong phase is approximately $SU(3)$ flavor symmetric.
\end{abstract}

%\newpage
\section{Introduction}
\setcounter{equation}{0}

To understand the quark mixing sector of the standard model(SM) and search
for new physics beyond the SM, one needs to study the decays of heavy mesons and
calculate precisely the transition matrix elements of the heavy mesons
decays. The short distance effects due to hard gluon exchange can be calculated
reliably and the effective hamiltonian and factorization approach has been
constructed \cite{effect,bsw}, thus a lot of results which fit the experimental
data well have been obtained  by factorization approach. However, there are
still many decay modes which can not be accommodated by factorization approach.
In fact, the quarks in heavy mesons are bound by strong interaction which
is described by nonperturbative $QCD$. After weak decays of heavy mesons,
the final particles can rescatter into other particle states through
nonperturbative strong interaction \cite{fsi, Donoghue}, this is called
final state interaction (FSI). Many authors have studied $FSI$ effects
and found that $FSI$ effects may play a crucial role in some decay
modes \cite{buccella,important}. Therefore it is necessary to study
heavy meson two-body weak decays beyond the factorization
approach. Since the $FSI$ process refers to the soft
rescattering process which is controlled by nonperturbative
$QCD$ and can not be reliably evaluated with well-established theoretical
frame, we have to  rely on phenomenological models to analyze the FSI effects
in certain processes. One can
model this rescattering effect as one-particle exchange process
\cite{ope1,ope2}. There are also other ways to treat the nonperturbative and
FSI effects in $D$ decays, the readers can refer to Ref.\cite{ref}.  In this
paper, we study some channels of $D\to PV$ decays. We use the
one-particle-exchange method to study the final state interactions in these
decays. The magnitudes of hadronic couplings needed here are extracted from
experimental data on the measured branching fractions of resonance decays. In
addition, we consider a strong phase for the hadronic
coupling \cite{duyang} which is important for obtaining the correct branching
ratios of $D\to PV$ decays. We also take into account some possible
nonfactorizable effect \cite{nonfactorization,cheng}, which is
needed for some decay mode from the phenomenological point of view.
The coupling constants extracted from experimental data are small for
$s$-channel contribution and large for $t$-channel contribution. Therefore the
$s$-channel contribution is numerically negligible in $D\to PV$ decays. We can
safely drop the $s$-channel contribution in this paper.

The paper is organized as follows. Section II presents the calculation
in naive factorization approach. Section III gives the main scheme of
one-particle-exchange method. Section IV is devoted to
the numerical calculation and discussions.
Finally a brief summary is given.

\section{Calculations in the factorization approach }

The low energy effective Hamiltonian for charm
decays is given by \cite{buras}
\begin{eqnarray}
{\cal H}_{eff}&=&\frac{G_F}{\sqrt2}\{V_{us}V_{cs}^*[C_1(\bar s c)_{V-A}(\bar u
s)_{V-A} +C_2(\bar s s)_{V-A}(\bar u c)_{V-A}]\; \nonumber \\ & &
+V_{ud}V_{cd}^*[C_1(\bar d c)_{V-A}(\bar u d)_{V-A}
+C_2(\bar d d)_{V-A}(\bar u c)_{V-A}]\; \nonumber \\ & &
+V_{ud}V_{cs}^*[C_1(\bar s c)_{V-A}(\bar u d)_{V-A}
+C_2(\bar s d)_{V-A}(\bar u c)_{V-A}]\}+h.c. \; ,
\label{Hamiltonian}
\end{eqnarray}
where $C_1$ and $C_2$ are the Wilson coefficients at $m_c$ scale.
We need not consider the contributions of the QCD and
electroweak penguin operators in the decays of $D \to PV$,
since their effects are small in $D$ decays. The values of $C_1$
and $C_2$ at $m_c$ scale are taken to be
$C_1=1.216$, $C_2=-0.415 $ \cite{buras}

In the naive factorization approach, the decay amplitude can be
generally factorized into a product of two current matrix elements
and can be obtained from eq.(\ref{Hamiltonian})
\begin{eqnarray}
A(D^0\to\bar{K^0}\rho^0)&=&\sqrt2\;G_F\;V_{ud}\;V_{cs}^*\;a_2\;
m_{\rho^0}\; f_{\pi}A^{D\rho}_0\;\epsilon_{\rho^0}\cdot
P_{\bar{k^0}},\nonumber\\
A(D^0\to K^-\rho^+)&=&\sqrt2\;G_F\;V_{ud}\;V_{cs}^*\;a_1\;m_{\rho^+}
    f_{\rho^+}F^{Dk}\;\epsilon_{\rho^+}\cdot P_{D^0},\nonumber\\
A(D^+\to\bar{K^0}\rho^+)&=&\sqrt2\;G_F\;V_{ud}\;V_{cs}^*\;
      m_{\rho^+}(a_1F_1^{Dk}\epsilon_{\rho^+}\cdot
P_{D^+}\;
+a_2f_K\;A_0^{D\rho^+}\;\epsilon_{\rho^+}\cdot P_{\bar{K^0}}),\nonumber\\
A(D^0\to\pi^0\bar{K^{*0}})&=&\sqrt2\;G_F\;V_{ud}\;V_{cs}^*\;m_{\bar{K^*_0}}\;
f_{\bar{K^*_0}}\; F_1^{D\pi}\;\epsilon_{\bar{K^*_0}}\cdot
P_{D^0},\nonumber\\
A(D^0\to \pi^+K^{*-})&=&\sqrt2\;G_F\;V_{ud}\;V_{cs}^*\;a_1\;
m_{K^*}f_{\pi}\;  A_0^{DK}\;\epsilon_{K^{*-}} \cdot P_{\pi^+}, \nonumber\\
A(D^+\to\pi^+\bar{K^{*0}})&=&\sqrt2\;G_F\;V_{ud}\;V_{cs}^*\;
   m_{\bar{K^*_0}}(a_1f_{\pi}A_0^{DK^*}\epsilon_{\bar{K^*_0}}\cdot P_{\pi^+}\;
   +a_2f_{\bar{K^*_0}}\;F^{D\pi}\;\epsilon_{\bar{K^*_0}}\cdot
P_{D^0}),\,\nonumber\\
A(D^+\to\pi^+\rho^0)&=&\sqrt2\;G_F\;V_{ud}\;V_{cd}^*\;a_1\;
m_{\rho^0}\;
f_\pi\;A_0^{D\rho}\;\epsilon_{\rho^0}\cdot P_\pi\;
-G_F\;V_{ud}\;V_{cd}^*\; a_2\;m_{\rho^0}\;f_\rho\;F_1^{D\pi}\;
\epsilon_\rho\cdot P_D, \,\nonumber\\
A(D^0\to \pi^+\rho^-)&=&\sqrt2\;G_F\;V_{ud}\;V_{cd}^*\;m_{\rho^-}\;
   f_\pi\;A_0^{D\rho}\;\epsilon_{\rho^-}\cdot P_\pi, \,\nonumber\\
A(D^0\to\pi^0\rho^0)&=&-G_F\;V_{ud}\;V_{cd}^*\;a_2\;m_{\rho^0}(f_\pi\;
f_\pi\;A_0^{D\rho}\;\epsilon_\rho\cdot P_\pi \;
+f_\rho\;F_1^{D\pi}\;\epsilon_{\rho^0} \cdot P_{D^0}),\nonumber\\
A(D^0\to\pi^-\rho^+)&=&\sqrt2\;G_F\;V_{ud}\;V_{cd}^*\;a_1\;m_{\rho^+}
   f_\rho\;F_1^{D\pi}\;\epsilon_\rho\cdot P_{D^0}; \,\nonumber\\
A(D^+\to\pi^0\rho^+)&=&\frac{G_F}{2}\;V_{ud}\;V_{cd}^* \;
m_{\rho}(2a_1\;     f_\rho\;F_1^{D\pi}\;\epsilon_{\rho^+}\cdot
 P_{D^0}   -\sqrt2\;
a_2\;f_\pi \;
A^{D\rho}\;\epsilon_{\rho^0}\cdot P_{\pi^0}) ,
\end{eqnarray}
where the parameters $a_1$ and $a_2$ are taken as \cite{cheng}
\begin{eqnarray}  \label{cheng}
a_1= c_1(\mu) + c_2(\mu) \left({1\over N_c} +\chi(\mu)\right)\,,
\qquad \quad a_2 = c_2(\mu) + c_1(\mu)\left({1\over N_c} +
\chi(\mu)\right) ,
\end{eqnarray}
with the color number $N_c=3$, and $\chi(\mu)$ is the phenomenological
parameter introduced for taking care of  nonfactorizable effects. The
parameters in
 calculation are:  1) the form factors, $F_1^{D\pi}(0)=0.69$,
$F_1^{DK}(0)=0.76$, $A_0^{D\rho}(0)=0.67 $,  $A_0^{DK^*}(0)=0.73 $
\cite{bsw}; 2) the decay constants, $f_{\pi}=0.133GeV$, $f_K=0.158GeV$,
$f_{\rho}=0.2GeV$, and $f_{K^*}=0.221GeV$.

For $q^2$ dependence of the form factors, we take
the BSW model \cite{bsw}, i.e., the monopole dominance assumption:
\begin{equation}
F_1(q^2)=\frac{F_1(0)}{1-q^2/m_{1^-}^2},\qquad
A_0(q^2)=\frac{A_0(0)}{1-q^2/m_{0^-}^2},
\end{equation}
where $m_{1^-},m_{0^-}$ is the relevant pole mass.

The decay width of a $D$ meson at rest decaying into $PV$ is
\begin{equation}
\Gamma(D \to PV)=\frac{1}{8\pi}|A(D \to PV)|^2\frac{|\vec
p\;|}{m_D^2},
\end{equation}
where the momentum of the final state particle is given by
\begin{equation}
  |\vec p\;|=\frac{[(m_D^2-(m_1+m_2)^2))
((m_D^2-(m_1-m_2)^2)]^{1/2}}{2m_D},
\end{equation}
where $m_1,m_2 $ are the masses of final state particles.
The corresponding branching ratio is
\begin{equation}
 Br(D \to PV)=\frac{\Gamma(D \to PV)}{\Gamma_{tot}}.
\end{equation}

\begin{table}[h]
\caption{{\small The branching ratios of $D \to PV $ obtained in the naive
factorization approach and compared with the experimental
results.}}
\begin{center}
\begin{tabular}{|c|c|c|c|} \hline
Decay mode & Br (Theory) &Br (Theory) & Br
(Experiment)\\
$ $& $(\chi=0)$ &($a_1=1.26,a_2=-0.51)$& \\ \hline
$D^0 \to \bar{K^0}\rho^0$ &$3.92 \times 10^{-3}$ &$5.92 \times 10^{-3}$&
$(1.21\pm 0.17)\times 10^{-2}$ \\ \hline
$D^0 \to K^-\rho^+$ &$10.66\times 10^{-2}$ &$11.45 \times 10^{-2}$&$(10.8\pm
0.9) \times 10^{-2}$\\
\hline
$D^+\to \bar{K^0}\rho^+$ &$17.35 \times 10^{-2}$  &$ 16.91 \times 10^{-2}$&
$(6.6 \pm 2.5) \times 10^{-2}$\\
\hline
$D^0 \to \pi^0\bar{K^*_0}$ &$1.37 \times 10^{-2}$ &$ 2.08 \times 10^{-2}$ &
$(3.1\pm 0.4)\times 10^{-2}$ \\ \hline
$D^0 \to \pi^+K^-$ &$3.06 \times 10^{-2}$ &$3.29 \times 10^{-2}$ & $(5.0\pm
0.4)\times 10^{-2}$ \\ \hline
$D^+ \to \pi^+\bar{K^*_0}$ &$8.72 \times 10^{-3}$ &$ 3.66 \times 10^{-2}$&
$(1.90\pm 0.19)\times 10^{-2}$ \\ \hline
$D^+ \to \pi^+\rho^0$ &$8.12 \times 10^{-3}$ &$ 9.33 \times 10^{-3}$&
$(1.05\pm 0.31) \times10^{-3}$\\ \hline
$D^0 \to \pi^+\rho^-$ &$1.36 \times 10^{-3}$ &$ 1.46 \times 10^{-3}$& $-$ \\
\hline $D^0 \to \pi^0\rho^0$ &$7.17 \times 10^{-4}$ &$ 1.09 \times 10^{-3}$&
$-$ \\ \hline $D^0 \to \pi^-\rho^+$ &$4.48 \times 10^{-3}$ & $ 5.89 \times
10^{-3}$&$-$ \\ \hline
$D^+ \to \pi^0\rho^+$ &$1.78 \times 10^{-2}$ &$ 1.98 \times
10^{-2}$ & $-$ \\  \hline
\end{tabular}
\end{center}
\end{table}

The numerical results of the branch ratios of $D$ decays are given in Table
1. The second column is for the case $\chi(\mu)=0$ which means
there is no nonfactorizable contribution. When
$\chi(\mu)=0$, the parameters $a_1 =1.216$ and $a_2=-0.415$.
The parameters in the third column  $a_1=1.26$,
$a_2=-0.51$ are phenomenologically used in many references \cite{phen}, which
is relevant to taking non-zero parameter $\chi(\mu)$.

Comparing the results of the naive factorization  in
the second and third column  of Table 1 with the experimental data,
one can notice that, even considering some nonfactorizable contribution,
some of the results from the naive
factorization approach deviate significantly from the experimental
data.

\section{The one particle exchange method for FSI}

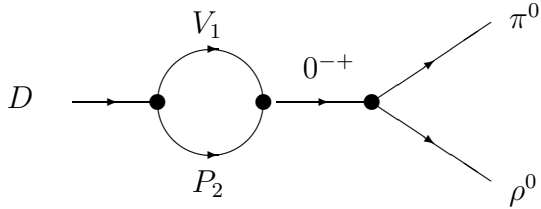
\begin{figure}[h]
\begin{center}
\begin{picture}(250,90)
      \put(35,50){\line(1,0){30}}
      \put(67,50){\circle*{6}}
      \put(87,50){\circle{80}}
      \put(107,50){\circle*{6}}
      \put(112,50){\line(1,0){38}}
      \put(148,50){\circle*{6}}
      \put(148,50){\line(3,-2){45}}
      \put(148,50){\line(3,2){45}}

   \put(50,50){\vector(3,0){2}}
   \put(88,70){\vector(3,0){2}}
   \put(88,30){\vector(3,0){2}}
   \put(130,50){\vector(3,0){2}}
   \put(170,64.5){\vector(3,2){2}}
   \put(170,35.5){\vector(3,-2){2}}

      \put(10,48){$D$}
      \put(80,16){$P_2$}
      \put(80,76){$V_1$}
      \put(122,59){$0^{-+}$}
      \put(200,78){$\pi^0$}
      \put(200,13){$\rho^0$}
   \end{picture}
 \end{center}
 \caption{{\small s-channel contributions to final-state interaction in $D\to
     PV$ decays.}}
 \label{schan}
\end{figure}

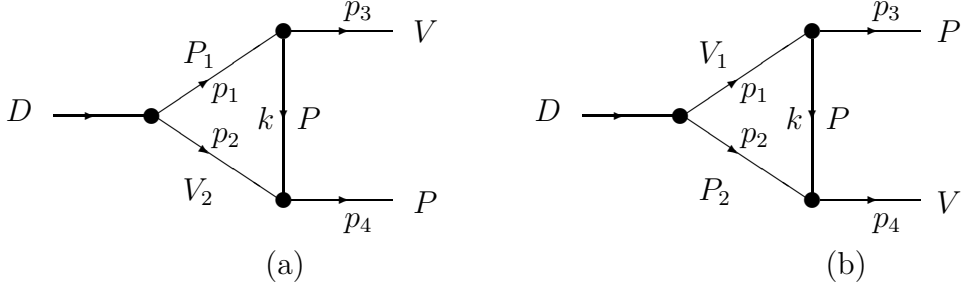
\begin{figure}[h] \begin{center}
\begin{picture}(430,160)
%------------------------(a)
\put(28,65){\line(1,0){37}}
\put(65,65){\circle*{6}}
\put(66,65){\line(3,2) {48}}
\put(66,65){\line(3,-2){48}}
\put(115,33){\circle*{6}}
\put(115,97){\circle*{6}}
\put(115,36){\line(0,1){58}}
\put(118,33){\line(1,0){38}}
\put(118,97){\line(1,0){38}}

\put(42,65){\vector(3,0){2}}
\put(85,52){\vector(3,-2){2}}
\put(85,77.5){\vector(3,2){2}}
\put(115,65){\vector(0,-2){2}}
\put(138,97){\vector(3,0){2}}
\put(138,33){\vector(3,0){2}}

\put(10,63){$D$}
\put(77,33){$V_2$}
\put(77,85){$P_1$}
\put(120,60){$P$}
\put(88,72){$p_1$}
\put(88,55){$p_2$}
\put(105,60){$k$}
\put(164,93){$V$}
\put(164,28){$P$}
\put(138,103){$p_3$}
\put(138,23){$p_4$}
\put(108,5){(a)}
%\end{picture}

%\begin{picture}(0,0)
%------------------------(b)
\put(228,65){\line(1,0){38}}
\put(265,65){\circle*{6}}
\put(266,65){\line(3,2) {48}}
\put(266,65){\line(3,-2){48}}
\put(315,33){\circle*{6}}
\put(315,97){\circle*{6}}
\put(315,36){\line(0,1){58}}
\put(318,33){\line(1,0){38}}
\put(318,97){\line(1,0){38}}

 \put(242,65){\vector(3,0){2}}
 \put(285,52){\vector(3,-2){2}}
 \put(285,77.5){\vector(3,2){2}}
 \put(315,65){\vector(0,-2){2}}
 \put(338,97){\vector(3,0){2}}
 \put(338,33){\vector(3,0){2}}

  \put(210,63){$D$}
  \put(272,33){$P_2$}
  \put(272,85){$V_1$}
  \put(320,60){$P$}
  \put(288,72){$p_1$}
  \put(288,55){$p_2$}
  \put(305,60){$k$}
  \put(362,93){$P$}
  \put(362,28){$V$}
  \put(338,103){$p_3$}
  \put(338,23){$p_4$}
  \put(320,5){(b)}
\end{picture}
\end{center}
\caption{{\small t-channel contributions to final-state interaction in $D\to
 PV$ due to one particle  exchange.
 (a) $D\to P_1V_2 \to PV $, (b) $D\to V_1P_2 \to PV $.}}
 \label{tchan}
\end{figure}

From Table 1, we can see that the calculation from naive
factorization approach is in disagreement with the experimental results for
the branching ratios of $D\to PV $ decays. The reason is that the
physical picture of naive factorization is too simple, in which
nonperturbative strong interaction is restricted in single hadrons, or
between the initial and final hadrons which share the same spectator quark.
If the mass of the initial particle is large, such as the case of $B$ meson
decay, the effect of nonperturbative strong interaction between the final
hadrons most probably is small because the momentum transfer is large.
However, in the case of $D$ meson, its mass is not so large. The energy scale
of $D$ decays is not very high. Nonperturbative effect may give large
contribution. According to the one-particle exchange method,
there are $s$-channel and $t$-channel contribution to the
final state interaction  \cite{ope1,ope2}. The diagrams of these
nonperturbative rescattering effects can be depicted in Figs.\ref{schan} and
\ref{tchan}. The first part $D \to P_1V_2$ or $D \to V_1P_2$ represents the
direct decay where the decay amplitudes can be obtained by using naive
factorization method. The second part represents rescattering process where
the effective hadronic couplings are needed in numerical calculation, which
can be extracted from experimental data on the relevant resonance decays.

There are many resonances near the mass scale of $D$ meson, it is
possible that nonperturbative interaction is propagated by these resonance
states, such as, $K^*(892)$, $K^*(1430)$,
$f_0(1710)$, $K^*(1680)$, $K^*(1020)$, $\phi(1680)$, $\pi(1300) $,  etc.
For s-channel the correct quantum number of the resonance should be
$J^{P}=0^{-}$ (in charged $D$ decays). In neutral decay modes, the reasonance
should be with quantum number $J^{PC}=0^{-+}$.
For $D^0\to\pi^0\rho^0$, only $\pi(1300)$ has the
correct quantum number \cite{PDG}. Fig.\ref{schan} is the  $s$-channel
contribution to the final state interaction in $D^0\to\pi^0\rho^0$.  Here
$V_1$ and $P_2$ are the intermediate mesons. Because the coupling of
$\pi(1300)$ with $\pi^0\rho^0$ is too small \cite{duyang}, we can ignore the
s-channel contribution in the numerical analysis.
Fig.\ref{tchan} shows the $t$-channel contribution to the final
state interaction. $P_1$, $V_2$ and $V_1$, $P_2$ are the intermediate
states from direct weak decays. They rescatter into the final
states by exchanging one resonance state $P$. In this paper the
intermediate states are treated to be on their mass shell, because their
off-shell contribution can be attributed to the quark level. We assume the
on-shell contribution dominates in the final state interaction.
The exchanged resonances are
treated as a virtual particle. Their propagators are taken as
Breit-Wigner form,
\begin{equation}
\frac{i}{k^2-m^2+im\Gamma_{tot}},
\end{equation}
where $\Gamma_{tot}$ is the total decay width of the exchanged resonance.

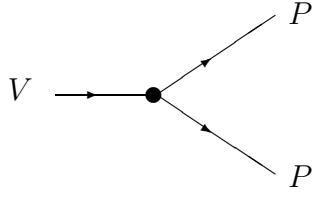
\begin{figure}[h]
\begin{center}
\begin{picture}(150,60)

\put(28,35){\line(1,0){38}}
\put(65,35){\circle*{6}}
\put(66,35){\line(3,2) {45}}
\put(66,35){\line(3,-2){45}}

\put(42,35){\vector(3,0){2}}
\put(85,22){\vector(3,-2){2}}
\put(85,47.5){\vector(3,2){2}}

\put(10,33){$V$}
\put(116,0){$P$}
\put(116,62){$P$}

\end{picture}
\end{center}
\caption{{\small The effective coupling vertex on the hadronic
level}}
\label{vertex1}
\end{figure}

We consider the $t$ channel contribution. For the $t$-channel contribution,
the concerned effective vertex in Fig.\ref{vertex1} is $VPP$, which can be
related to the $V$ decay amplitude. Explicitly the amplitude of $V\to PP$ can
be written as \begin{equation}
T_{VPP}=g_{VPP}\;\epsilon\cdot (p_1-p_2),
\label{t2}
\end{equation}
where $p_1$ and $p_2$ are the four-momentum of the two pseudoscalars,
respectively. To extract $g_{VPP}$ from experiment,
one should square eq.(\ref{t2}) to get the decay widths,
\begin{eqnarray}
\Gamma(V\to PP)&=& \frac{1}{3}\frac{1}{8\pi}\mid g_{VPP}\mid^2
         \left[m_V^2-2m_1^2-2m_2^2+\frac{(m_1^2-m_2^2)^2}{m_V^2}\right]
          \frac{\mid\vec{p}\mid}{m_V^2},
\label{couple}
\end{eqnarray}

\noindent where $m_1$ and $m_2$ are the masses of the two final particles $PP$,
respectively, and $\mid\vec{p}\mid$ is the momentum of one of the final
particle $P$ in the rest frame of $V$. From the above equations,
one can see that only the magnitudes of the effective couplings
 $\mid g_{VPP}\mid$ can be extracted from
experiment. On the quark level, the effective vertex should be controlled
by nonperturbative $QCD$. It is reasonable that a strong phase can appear in
the effective coupling, which is contributed by strong interaction.
Therefore we can take  a strong phase for each
hadronic effective
coupling  \cite {duyang}. In the following, the symbol
$g$ will only be used to represent the magnitude of the relevant
effective coupling. The total one should be $g e^{i \theta}$, where
$\theta$ is the strong phase. For example,
the effective couplings will be written in the form of
$g_{VPP}e^{i\theta_{VPP}}$.

The $t$-channel contribution in  Fig.\ref{tchan}(a) is
\begin{eqnarray}
  A^{FSI}_{P_1,V_2}&=&\frac12 \int \frac{d^3 \vec p_1}{(2\pi)^3 2E_1}
          \int\frac{d^3 \vec p_2}{(2\pi)^3 2E_2} (2\pi)^4
 \delta^4(p_D-p_1-p_2) A(D \to P_1V_2)\nonumber\\
  & &\hspace*{1cm}\times g_1 \; \epsilon_3\cdot (p_1+k)
  \frac{i\;e^{i(\theta_1+\theta_2)}}{k^2-m^2+im\Gamma_{tot}} \;
 F(k^2)^2 \;g_2 \; \epsilon_2\cdot(p_4+k),
\label{fsisk}
\end{eqnarray}
where
$F(k^2)=(\Lambda^2-m^2)/(\Lambda^2-k^2)$
is the
form factor which is introduced to compensate the off-shell effect of the
exchanged particle at the vertices \cite{Fk}.  We choose the lightest
resonance state as the exchanged particle that gives  the largest
contribution to the decay amplitude. After a few steps of
integration to the above equation, we get
\begin{equation}
  A^{FSI}_{P_1,V_2}=\int^1_{-1}  \frac{d(\cos\theta)}{2\pi m_D} |\vec p_1|
   X_1\;g_1 \frac{i\;e^{i(\theta_1+\theta_2)}}
  {k^2-m^2+im\Gamma_{tot}} \;F(k^2)^2\;g_2\;H_1\; ,
\label{fsitv}
\end{equation}
where
\begin{eqnarray}
  H_1&=&m_2\;f_1\;A_0[-(E_1E_4+|\vec p_1||\vec p_4|\cos\theta)\nonumber\\
& &+\frac{1}{2m_2^2}(M_D^2-m_1^2-m_2^2)(E_2E_4-|\vec p_2|
    |\vec p_4|\cos\theta)] \nonumber \\ & &
\times[\frac{1}{m_3}(|\vec p_3|E_1-E_3|\vec p_1|)],
\end{eqnarray}
and $X_1$ represents the relevant direct decay amplitude of $D$ decaying
to the intermediate pair $P_1$ and $V_2$ divided by
$\langle P_1 |(V-A)_\mu |0\rangle \langle V_2|(V-A)^\mu|D\rangle$,
$$
X_1\equiv \frac{A(D\to P_1 V_2)}
  {\langle P_1 |(V-A)_\mu |0\rangle \langle
  V_2|(V-A)^\mu|D\rangle}.
$$

The $t$-channel contribution in  Fig.\ref{tchan}(b) is
\begin{eqnarray}
  A^{FSI}_{V_1,P_2}&=&\frac12 \int \frac{d^3 \vec p_1}{(2\pi)^3 2E_1}
          \int\frac{d^3 \vec p_2}{(2\pi)^3 2E_2} (2\pi)^4
 \delta^4(p_D-p_1-p_2) A(D \to V_1P_2)\nonumber\\
  & &\hspace*{1cm} \times g_1 \; \epsilon_1 \cdot (p_3-k)
  \frac{i\;e^{i(\theta_1+\theta_2)}}{k^2-m^2+im\Gamma_{tot}} \; F(k^2)^2
    \;g_2 \; \epsilon_4\cdot(p_2-k),
\end{eqnarray}
and we obtain
\begin{eqnarray}
 A^{FSI}_{V_1,P_2}=\int^1_{-1} \frac{d(\cos\theta)}{2\pi m_D} |\vec p_1| \;
    \frac{i\;e^{i(\theta_1+\theta_2)}}{k^2-m^2+im\Gamma_{tot}}
     X_2\;  g_1 \;g_2 \; F(k^2)^2\;H_2\; ,
\label{fsitp}
\end{eqnarray}
where
\begin{eqnarray}
H_2&=&m_1f_1F_1[-M_DE_3+\frac{1}{m_1^2}\;E_1M_D(E_1E_3-|\vec p_1||\vec
p_3|\cos\theta)] \nonumber \\ & & \times\frac{1}{m_4}
\; (|{\vec p_4}|E_2-E_4|\vec p_2|\cos\theta),
\end{eqnarray}
and $X_2$ represents the relevant direct decay amplitude of $D$ decaying
to the intermediate  pair $V_1$ and $P_2$ divided by
$\langle V_1 |(V-A)_\mu |0\rangle \langle P_2|(V-A)^\mu|D\rangle$,
$$
X_2\equiv \frac{A(D\to V_1 P_2)}
  {\langle V_1 |(V-A)_\mu |0\rangle \langle
  P_2|(V-A)^\mu|D\rangle}.
$$

\section{Numerical calculation and  discussions}
To calculate $FSI$ contribution of $D$ decays with
the eq.(\ref{fsitv}) and eq.(\ref{fsitp}), we need to know which channels can
rescatter into the final states. For $D\to K\rho$, $ \pi K^*$, $\pi\rho$, from
Figs.\ref{r1}$\sim$\ref{r3}, one can see that $D \to\pi K^* \to K \rho$,
$D \to  K \rho \to K \rho$, $D \to \pi K^* \to \pi K^*$,
$D \to\rho K \to \pi K^*$, $D \to \pi\rho \to \pi\rho$,
$D \to K K^* \to\pi\rho$ can give the largest  contributions,
because these intermediate states have the largest  couplings
with the final states and the  masses of exchanged
meson are small which  give the largest $t$-channel contributions.
When calculating the contribution of each diagram in Figs.\ref{r1}$\sim$\ref{r3},
we should, at first, consider all the possible iso-spin
structure for each diagram  in Figs.\ref{r1}$\sim$\ref{r3}, and draw
all the possible sub-diagrams on the quark
level. Second, write down the iso-spin factor for each sub-diagram. For
example, the $u\bar{u}$ component in one final meson $\rho^0$ contributes an
isospin factor $\frac{1}{\sqrt{2}}$, and the $d\bar{d}$ component contributes
$-\frac{1}{\sqrt{2}}$. For the intermediate state $\rho^0$, the factor
$\frac{1}{\sqrt{2}}$ and $-\frac{1}{\sqrt{2}}$ should be dropped
\cite{duyang}. Third, sum the factors of all the possible sub-diagrams
of each diagram to get the iso-spin factor for each diagram on the hadronic
level. For example, in the diagram (a) of  $D^0\to\bar{K^0}\rho^0 $, the
iso-spin factor of one sub-diagram is $\frac{1}{\sqrt{2}}$ and is
$-\frac{1}{\sqrt{2}}$ in another sub-diagram, so the  factor of the
diagram (a) of $D^0\to\bar{K^0}\rho^0 $ is zero.

\begin{figure}[h]
\setlength{\unitlength}{0.012in}
\begin{center}
\begin{picture}(430,100)
%------------------------(a)
\put(28,65){\line(1,0){37}}
\put(65,65){\circle*{6}}
\put(66,65){\line(3,2) {48}}
\put(66,65){\line(3,-2){48}}
\put(115,33){\circle*{6}}
\put(115,97){\circle*{6}}
\put(115,36){\line(0,1){58}}
\put(118,33){\line(1,0){38}}
\put(118,97){\line(1,0){38}}

\put(42,65){\vector(3,0){2}}
\put(85,52){\vector(3,-2){2}}
\put(85,77.5){\vector(3,2){2}}
\put(115,65){\vector(0,-2){2}}
\put(138,97){\vector(3,0){2}}
\put(138,33){\vector(3,0){2}}

\put(10,63){$D$}
\put(77,33){$K^*$}
\put(77,85){$\pi$}
\put(120,60){$\pi$}
\put(164,93){$\rho$}
\put(164,28){$K$}
\put(108,5){(a)}

%------------------------(b)
\put(228,65){\line(1,0){37}}
\put(265,65){\circle*{6}}
\put(266,65){\line(3,2) {48}}
\put(266,65){\line(3,-2){48}}
\put(315,33){\circle*{6}}
\put(315,97){\circle*{6}}
\put(315,36){\line(0,1){58}}
\put(318,33){\line(1,0){38}}
\put(318,97){\line(1,0){38}}

 \put(242,65){\vector(3,0){2}}
 \put(285,52){\vector(3,-2){2}}
 \put(285,77.5){\vector(3,2){2}}
 \put(315,65){\vector(0,-2){2}}
 \put(338,97){\vector(3,0){2}}
 \put(338,33){\vector(3,0){2}}

  \put(210,63){$D$}
  \put(272,33){$\pi$}
  \put(272,85){$K^*$}
  \put(320,60){$\pi$}
  \put(362,93){$K$}
  \put(362,28){$\rho$}
  \put(320,5){(b)}
\end{picture}
\end{center}

\begin{center}
\begin{picture}(430,100)
%------------------------(c)
\put(28,65){\line(1,0){37}}
\put(65,65){\circle*{6}}
\put(66,65){\line(3,2) {48}}
\put(66,65){\line(3,-2){48}}
\put(115,33){\circle*{6}}
\put(115,97){\circle*{6}}
\put(115,36){\line(0,1){58}}
\put(118,33){\line(1,0){38}}
\put(118,97){\line(1,0){38}}

\put(42,65){\vector(3,0){2}}
\put(85,52){\vector(3,-2){2}}
\put(85,77.5){\vector(3,2){2}}
\put(115,65){\vector(0,-2){2}}
\put(138,97){\vector(3,0){2}}
\put(138,33){\vector(3,0){2}}

\put(10,63){$D$}
\put(77,33){$\rho$}
\put(77,85){$K$}
\put(120,60){$K$}
\put(164,93){$\rho$}
\put(164,28){$K$}
\put(108,5){(c)}

%------------------------(d)
\put(228,65){\line(1,0){37}}
\put(265,65){\circle*{6}}
\put(266,65){\line(3,2) {48}}
\put(266,65){\line(3,-2){48}}
\put(315,33){\circle*{6}}
\put(315,97){\circle*{6}}
\put(315,36){\line(0,1){58}}
\put(318,33){\line(1,0){38}}
\put(318,97){\line(1,0){38}}

 \put(242,65){\vector(3,0){2}}
 \put(285,52){\vector(3,-2){2}}
 \put(285,77.5){\vector(3,2){2}}
 \put(315,65){\vector(0,-2){2}}
 \put(338,97){\vector(3,0){2}}
 \put(338,33){\vector(3,0){2}}

  \put(210,63){$D$}
  \put(272,33){$K$}
  \put(272,85){$\rho$}
  \put(320,60){$K$}
  \put(362,93){$K$}
  \put(362,28){$\rho$}
  \put(320,5){(d)}
%  \put(215,0){(I)}
\end{picture}
\end{center}
\caption{{\small Intermediate states in rescattering process
for $D\to K\rho$ decays.}} \label{r1}
\end{figure}
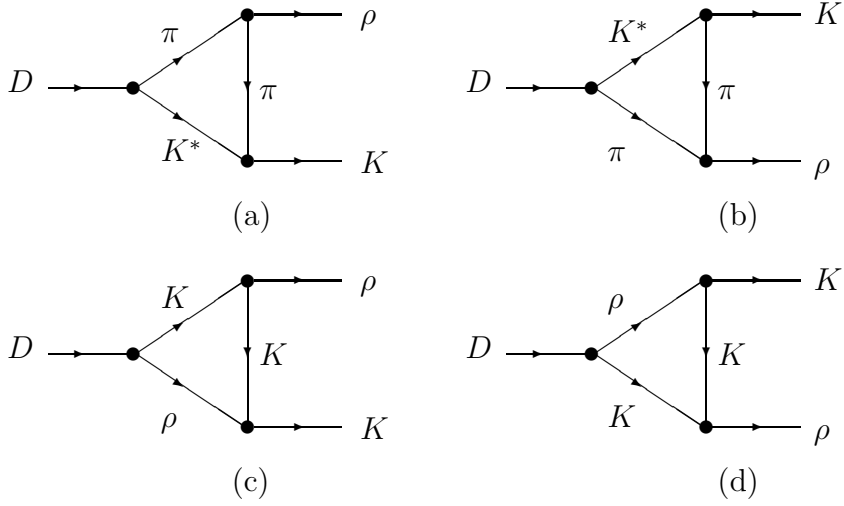
\begin{figure}[h]
\setlength{\unitlength}{0.012in}
\begin{center}
\begin{picture}(430,100)
%------------------------(a)
\put(28,65){\line(1,0){37}}
\put(65,65){\circle*{6}}
\put(66,65){\line(3,2) {48}}
\put(66,65){\line(3,-2){48}}
\put(115,33){\circle*{6}}
\put(115,97){\circle*{6}}
\put(115,36){\line(0,1){58}}
\put(118,33){\line(1,0){38}}
\put(118,97){\line(1,0){38}}

\put(42,65){\vector(3,0){2}}
\put(85,52){\vector(3,-2){2}}
\put(85,77.5){\vector(3,2){2}}
\put(115,65){\vector(0,-2){2}}
\put(138,97){\vector(3,0){2}}
\put(138,33){\vector(3,0){2}}

\put(10,63){$D$}
\put(77,33){$K^*$}
\put(77,85){$\pi$}
\put(120,60){$K$}
\put(164,93){$K^*$}
\put(164,28){$\pi$}
\put(108,5){(a)}

%------------------------(b)
\put(228,65){\line(1,0){37}}
\put(265,65){\circle*{6}}
\put(266,65){\line(3,2) {48}}
\put(266,65){\line(3,-2){48}}
\put(315,33){\circle*{6}}
\put(315,97){\circle*{6}}
\put(315,36){\line(0,1){58}}
\put(318,33){\line(1,0){38}}
\put(318,97){\line(1,0){38}}

 \put(242,65){\vector(3,0){2}}
 \put(285,52){\vector(3,-2){2}}
 \put(285,77.5){\vector(3,2){2}}
 \put(315,65){\vector(0,-2){2}}
 \put(338,97){\vector(3,0){2}}
 \put(338,33){\vector(3,0){2}}

  \put(210,63){$D$}
  \put(272,33){$\pi$}
  \put(272,85){$K^*$}
  \put(320,60){$K$}
  \put(362,93){$\pi$}
  \put(362,28){$K^*$}
  \put(320,5){(b)}
\end{picture}
\end{center}

\begin{center}
\begin{picture}(430,100)
%------------------------(c)
\put(28,65){\line(1,0){37}}
\put(65,65){\circle*{6}}
\put(66,65){\line(3,2) {48}}
\put(66,65){\line(3,-2){48}}
\put(115,33){\circle*{6}}
\put(115,97){\circle*{6}}
\put(115,36){\line(0,1){58}}
\put(118,33){\line(1,0){38}}
\put(118,97){\line(1,0){38}}

\put(42,65){\vector(3,0){2}}
\put(85,52){\vector(3,-2){2}}
\put(85,77.5){\vector(3,2){2}}
\put(115,65){\vector(0,-2){2}}
\put(138,97){\vector(3,0){2}}
\put(138,33){\vector(3,0){2}}

\put(10,63){$D$}
\put(77,33){$\rho$}
\put(77,85){$K$}
\put(120,60){$\pi$}
\put(164,93){$K^*$}
\put(164,28){$\pi$}
\put(108,5){(c)}

%------------------------(d)
\put(228,65){\line(1,0){37}}
\put(265,65){\circle*{6}}
\put(266,65){\line(3,2) {48}}
\put(266,65){\line(3,-2){48}}
\put(315,33){\circle*{6}}
\put(315,97){\circle*{6}}
\put(315,36){\line(0,1){58}}
\put(318,33){\line(1,0){38}}
\put(318,97){\line(1,0){38}}

 \put(242,65){\vector(3,0){2}}
 \put(285,52){\vector(3,-2){2}}
 \put(285,77.5){\vector(3,2){2}}
 \put(315,65){\vector(0,-2){2}}
 \put(338,97){\vector(3,0){2}}
 \put(338,33){\vector(3,0){2}}

  \put(210,63){$D$}
  \put(272,33){$K$}
  \put(272,85){$\rho$}
  \put(320,60){$\pi$}
  \put(362,93){$\pi$}
  \put(362,28){$K^*$}
  \put(320,5){(d)}
%  \put(215,0) {(II)}
\end{picture}
\end{center}
\caption{{\small Intermediate states in rescattering process
for $D\to \pi K$ decays.}} \label{r2}
\end{figure}
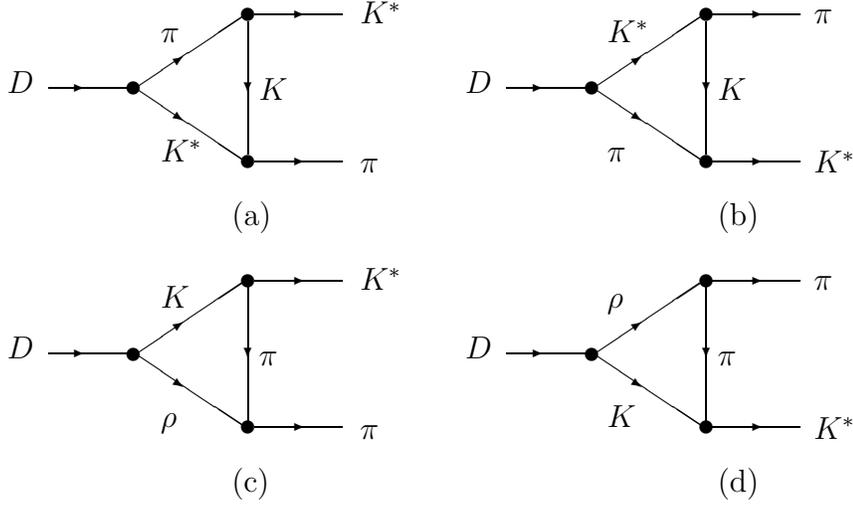
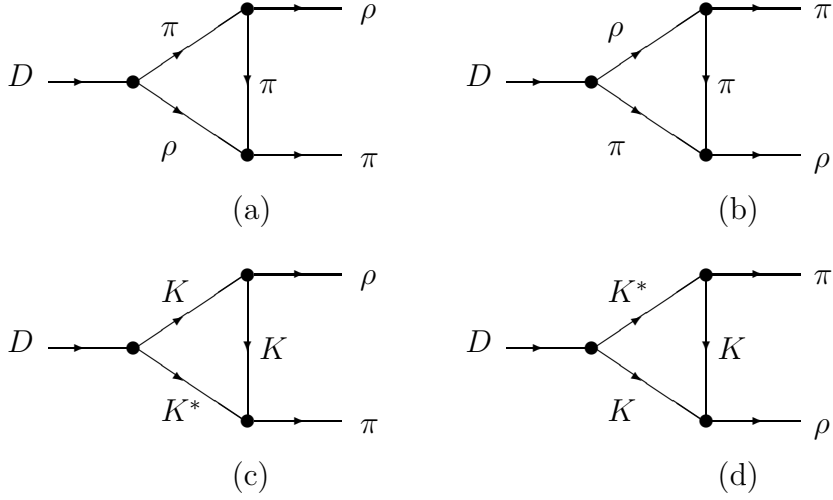
\begin{figure}[h]
\setlength{\unitlength}{0.012in}
\begin{center}
\begin{picture}(430,100)
%------------------------(a)
\put(28,65){\line(1,0){37}}
\put(65,65){\circle*{6}}
\put(66,65){\line(3,2) {48}}
\put(66,65){\line(3,-2){48}}
\put(115,33){\circle*{6}}
\put(115,97){\circle*{6}}
\put(115,36){\line(0,1){58}}
\put(118,33){\line(1,0){38}}
\put(118,97){\line(1,0){38}}

\put(42,65){\vector(3,0){2}}
\put(85,52){\vector(3,-2){2}}
\put(85,77.5){\vector(3,2){2}}
\put(115,65){\vector(0,-2){2}}
\put(138,97){\vector(3,0){2}}
\put(138,33){\vector(3,0){2}}

\put(10,63){$D$}
\put(77,33){$\rho$}
\put(77,85){$\pi$}
\put(120,60){$\pi$}
\put(164,93){$\rho$}
\put(164,28){$\pi$}
\put(108,5){(a)}

%------------------------(b)
\put(228,65){\line(1,0){37}}
\put(265,65){\circle*{6}}
\put(266,65){\line(3,2) {48}}
\put(266,65){\line(3,-2){48}}
\put(315,33){\circle*{6}}
\put(315,97){\circle*{6}}
\put(315,36){\line(0,1){58}}
\put(318,33){\line(1,0){38}}
\put(318,97){\line(1,0){38}}

 \put(242,65){\vector(3,0){2}}
 \put(285,52){\vector(3,-2){2}}
 \put(285,77.5){\vector(3,2){2}}
 \put(315,65){\vector(0,-2){2}}
 \put(338,97){\vector(3,0){2}}
 \put(338,33){\vector(3,0){2}}

  \put(210,63){$D$}
  \put(272,33){$\pi$}
  \put(272,85){$\rho$}
  \put(320,60){$\pi$}
  \put(362,93){$\pi$}
  \put(362,28){$\rho$}
  \put(320,5){(b)}
\end{picture}
\end{center}

\begin{center}
\begin{picture}(430,100)
%------------------------(c)
\put(28,65){\line(1,0){37}}
\put(65,65){\circle*{6}}
\put(66,65){\line(3,2) {48}}
\put(66,65){\line(3,-2){48}}
\put(115,33){\circle*{6}}
\put(115,97){\circle*{6}}
\put(115,36){\line(0,1){58}}
\put(118,33){\line(1,0){38}}
\put(118,97){\line(1,0){38}}

\put(42,65){\vector(3,0){2}}
\put(85,52){\vector(3,-2){2}}
\put(85,77.5){\vector(3,2){2}}
\put(115,65){\vector(0,-2){2}}
\put(138,97){\vector(3,0){2}}
\put(138,33){\vector(3,0){2}}

\put(10,63){$D$}
\put(77,33){$K^*$}
\put(77,85){$K$}
\put(120,60){$K$}
\put(164,93){$\rho$}
\put(164,28){$\pi$}
\put(108,5){(c)}

%------------------------(d)
\put(228,65){\line(1,0){37}}
\put(265,65){\circle*{6}}
\put(266,65){\line(3,2) {48}}
\put(266,65){\line(3,-2){48}}
\put(315,33){\circle*{6}}
\put(315,97){\circle*{6}}
\put(315,36){\line(0,1){58}}
\put(318,33){\line(1,0){38}}
\put(318,97){\line(1,0){38}}

 \put(242,65){\vector(3,0){2}}
 \put(285,52){\vector(3,-2){2}}
 \put(285,77.5){\vector(3,2){2}}
 \put(315,65){\vector(0,-2){2}}
 \put(338,97){\vector(3,0){2}}
 \put(338,33){\vector(3,0){2}}

  \put(210,63){$D$}
  \put(272,33){$K$}
  \put(272,85){$K^*$}
  \put(320,60){$K$}
  \put(362,93){$\pi$}
  \put(362,28){$\rho$}
  \put(320,5){(d)}
%  \put(215,0){(III)}
\end{picture}
\end{center}
\caption{{\small Intermediate states in rescattering process
for $D\to \pi\rho$ decays.}} \label{r3}
\end{figure}

%\caption{{\small Intermediate states in rescattering process
%for: (I): $D\to K\rho$ decays; (II): $D\to \pi K$ decays; (III):
%$D\to \pi\rho$ decays}} \label{r}
%\end{figure}

From eq.(\ref{fsitv}), eq.(\ref{fsitp}) and considering
Figs.\ref{r1}$\sim$\ref{r3}, we can calculate the amplitudes of
$D\to PV$ decays. In this paper we consider $D\to K\rho$, $\pi K^*$
and $\pi\rho$ decays. There should be some input parameters in our
calculation, such as, the transition form factors for $D$ decays,
decay constants of the final mesons, the phenomenological
nonfactorizable parameter $\chi (\mu)$, the off-shell
compensating parameter $\Lambda$ in function $F(k^2)$ introduced
in eq.(\ref{fsisk}), the effective couplings
of relevant hadronic states and the relevant strong phases for
these effective couplings. For the transition form factors and
decay constants, we take 1) the form factors, $F_1^{D\pi}(0)=0.69$,
$F_1^{DK}(0)=0.76$, $A_0^{D\rho}(0)=0.67 $,  $A_0^{DK^*}(0)=0.73 $
\cite{bsw}; 2) the decay constants, $f_{\pi}=0.133GeV$, $f_K=0.158GeV$,
$f_{\rho}=0.2GeV$, and $f_{K^*}=0.221GeV$. We should be careful
for these parameters, because except for the decay constants
$f_{\pi}$ and $f_K$, etc., the values of the transition form
factors have not been known exactly yet. We have to take them from
model-dependent calculations. For the phenomenological
nonfactorizable parameter $\chi (\mu)$, at first, we tried
to proceed by taking $\chi (\mu)=0$, which means that
nonfactorizable contribution is neglected. We find that
if nonfactorizable contribution is neglected, no matter
how the other parameters ( the strong phases and $\Lambda$ ) are
tuned, we can not reproduce the experimental data for all the
$D\to PV$ decays simultaneously. So we have to keep it as an
phenomenological parameter which will be determined later.
The hadronic effective couplings involved in this study are
$g_{\rho\pi\pi}$ and $g_{K^*K\pi}$, which can be determined from
the center value of the measured decay width of $\rho\to\pi\pi$
and $K^*\to K\pi$ \cite{PDG}. We obtain $g_{\rho\pi\pi}=6.0$,
$g_{K^*K\pi}=4.6$. The parameter $\Lambda$ in the off-shellness
compensating function $F(k^2)$ introduced in eq.(\ref{fsisk})
is not an universal parameter. It is process-dependent in general.
However, in this paper we use one value for $\Lambda$ in all the possible
channels of $D\to PV$ decays. We assume $\Lambda$ is in the range
from $0.5GeV$ to $1.0GeV$, which is the range of the masses of the
final state particles, $\rho$ and $K^*$, etc.. We scanned all the
possible value for $\chi(\mu)$ and $\Lambda$, and find that if we
take $\chi(\mu)=0.16$ and $\Lambda=0.7 GeV$, we can reproduce the
experimental data of all the detected $D\to PV$ decay modes well.
$\chi(\mu)=0.16$ means the nonfactorizable contribution is not
large. $\Lambda=0.7 GeV$ is in the mass range of the final
state particles. In the following we give the decay amplitudes
of some $D\to PV$ decay modes as function of the strong phases
$\theta_{K^*K\pi}$, $\theta_{\rho\pi\pi}$ and $\theta_{\rho KK}$
by taking $\chi(\mu)=0.16$ and $\Lambda=0.7GeV$,
\begin{eqnarray}
A(D^0\to\bar{K^0}\rho^0)&=&-6.768\times 10^{-7}+1.40508\times
10^{-6}i
 e^{i2\theta_{\rho KK}}, \nonumber\\
A(D^0\to K^-\rho^+)&=& 4.09 \times  10^{-6}-2.2512 \times 10^{-7}i
e^{i2\theta_{\rho KK}} -6.739\times 10^{-7}ie^{i(\theta_{K^*K\pi}+
\theta_{\rho\pi\pi})},\nonumber\\
A(D^+\to\bar{K^0}\rho^+)&=&3.2579\times  10^{-6}+1.14838\times  10^{-6}
ie^{i2\theta_{\rho KK}}+6.114\times 10^{-7}ie^{i(\theta_{K^*K\pi}+
\theta_{\rho\pi\pi})}, \nonumber\\
A(D^0\to\pi^0\bar{K^*_0})&=&-1.1837\times  10^{-6}+1.128\times  10^{-6}i
e^{i2\theta_{K^*K\pi}},\nonumber\\
A(D^0\to\pi^+K^{*-})&=&2.352\times  10^{-6}-5.501\times 10^{-7}i
e^{i2\theta_{K^*K\pi}}-4.51\times 10^{-7}ie^{i(\theta_{K^*K\pi}+
\theta_{\rho\pi\pi})}, \nonumber\\
A(D^+\to\pi^+\bar{K^*_0})&=&1.2056\times  10^{-6}+3.7368\times 10^{-7}i
e^{i2\theta_{K^*K\pi}}+1.81603\times  10^{-6}ie^{i(\theta_{K^*K\pi}+
\theta_{\rho\pi\pi})}, \nonumber\\
A(D^+\to\pi^+\rho^0)&=&-6.728887\times 10^{-7}-9.61\times 10^{-7}i
e^{i2\theta_{\rho KK}}+2.2334286\times 10^{-7}ie^{i(\theta_{K^*K\pi}+
\theta_{\rho KK})}, \nonumber\\
A(D^0\to\pi^+\rho^-)&=&-4.8216\times 10^{-7}+1.137\times 10^{-7}i
e^{i(\theta_{K^*K\pi}+\theta_{\rho KK})}+2.1\times 10^{-7}i
e^{i2\theta_{\rho\pi\pi}}, \nonumber\\
A(D^0\to\pi^0\rho^0)&=&-2.608\times 10^{-7}+1.821\times 10^{-7}i
e^{i(\theta_{K^*K\pi}+\theta_{\rho KK})}, \nonumber\\
A(D^0\to\pi^-\rho^+)&=&-8.736\times 10^{-7}+6.879\times 10^{-8}i
e^{i(\theta_{K^*K\pi}+\theta_{\rho KK})}+2.08\times 10^{-7}i
e^{i2\theta_{\rho\pi\pi}},\nonumber\\
A(D^+\to\pi^0\rho^+)&=&-1.0597\times  10^{-6}-9.6106\times 10^{-7}i
e^{i2\theta_{\rho\pi\pi}}+2.21\times 10^{-7}ie^{i(\theta_{K^*K\pi}+
\theta_{\rho\pi\pi})}.
\end{eqnarray}

The phases of the effective hadronic couplings $\theta_{K^*K\pi}$,
$\theta_{\rho\pi\pi}$ and $\theta_{\rho KK}$ can not be known
from direct experimental measurement or from any
nonperturbative calculations because
there are no any such kind of calculations yet. We only know
that the values of $\theta_{K^*K\pi}$, $\theta_{\rho\pi\pi}$
and $\theta_{\rho KK}$ should not differ too much
according to $SU(3)$ flavor symmetry. We tried some values
for these phase parameters,
and find that the ranges which can reproduce the experimental
data of the measured $D\to PV$ decays are not very narrow.
To show the situation that the experimental data are accommodated,
we give the numerical results for $\theta_{K^*K\pi}=51.0^\circ$,
$\theta_{\rho KK}=51.0^\circ$ and $\theta_{\rho\pi\pi}=57.3^\circ$ in table 2.

\begin{table}[h]
\caption{{\small The branching ratios of $D \to PV$.}}
\begin{center}
\begin{tabular}{|c|c|c|c|} \hline
Decay mode & Factorization & Factorization + FSI & Experiment\\
\hline
$D^0\to\bar{K^0}\rho^0$ &$3.92\times 10^{-3}$&
$1.25\times 10^{-2}$&$(1.21\pm 0.17)\times 10^{-2}$ \\
\hline
$D^0 \to K^-\rho^+$ &$10.66\times 10^{-2}$&
$11.1\times 10^{-2}$ &$(10.8 \pm 0.9) \times 10^{-2}$\\
\hline
$D^+\to\bar{K^0}\rho^+$ &$17.35\times 10^{-2}$ &
$7.01\times 10^{-2}$& $(6.6 \pm 2.5) \times 10^{-2}$\\
\hline
$D^0\to\pi^0\bar{K^*_0}$ &$1.37 \times 10^{-2}$ &
$2.72\times 10^{-2}$& $(3.1 \pm 0.4) \times 10^{-2}$\\
\hline
$D^0\to\pi^+K^-$ &$3.06 \times 10^{-2}$ &
$5.22\times 10^{-2}$& $(5.0 \pm 0.4) \times 10^{-2}$\\
\hline
$D^+\to\pi^+\bar{K^*_0}$ &$8.72 \times 10^{-3}$ &
$1.93\times 10^{-2}$& $(1.90 \pm 0.19) \times 10^{-2}$\\
\hline
$D^+\to\pi^+\rho^0$ &$8.12 \times 10^{-3}$ &
$1.3\times 10^{-3}$& $(1.05 \pm 0.31) \times 10^{-3}$\\
\hline
$D^0\to\pi^+\rho^-$ &$1.36 \times 10^{-3}$ &
$3.6\times 10^{-3}$&$-$\\
\hline
$D^0\to\pi^0\rho^0$ &$7.17 \times 10^{-4}$ &
$1.1\times 10^{-3}$& $-$\\
\hline
$D^0\to\pi^-\rho^+$ &$4.48 \times 10^{-3}$ &
$7.3\times 10^{-3}$& $-$\\
\hline
$D^+\to\pi^0\rho^+$ &$1.78 \times 10^{-2}$ &
$3.1\times 10^{-3}$& $-$\\
\hline
\end{tabular}
\end{center}
\label{tabbr2}
\end{table}

Table 2 shows that the contribution of $FSI$ is strongly channel
dependent. For example, For $D^0\to\bar{K^0}\rho^0 $,
the braching ratio in naive factorization is $3.922\times 10^{-3}$,
while the braching ratio including $FSI $ is
$1.25\times 10^{-2}$. we can see that FSI contribution in
$D^0\to\bar{K^0}\rho^0$ is large, but $FSI$ contribution in
$ D^0\to K^-\rho^+$ is small. The reason for the difference is
that the external rescattering diagrams for $D^0\to\bar{K^0}\rho^0$
and $D^0\to K^-\rho^+$ are different. Without the contribution
of $FSI$, predictions of naive factorization for most detected
$D\to PV$ decays are seriously in disagreement with the
experimental results. After including $FSI$, the results can
accommodate the experimental data well. For the other decay modes
$D^0\to \pi^+\rho^-$, $\pi^0\rho^0$, $\pi^-\rho^+$ and
$D^+\to \pi^0\rho^+$, their branching ratios have not been
detected in experiment yet. In our model, they are all predicted to
be at the order of ${\cal O}(10^{-3})$. For $D^0\to \pi^+\rho^-$
and $\pi^-\rho^+$, the effect of $FSI$ is enhancement. While for
$D^0\to\pi^0\rho^0$ and $D^+\to \pi^0\rho^+$, $FSI$ suppresses
the prediction of naive factorization.

Before the end of this section, some comments should be given:
there are many uncertainties in the input parameters which may change
the above result numerically, such as, the $D$ decay transition
form factors and some decay constants which have not been known exactly
yet, they need to be measured from leptonic and semileptonic decays
of $D$ mesons which are quite possible in CLEO-C program in the
near future. The other sources which may cause uncertainties are
the shape of the off-shell compensating function $F(k^2)$, or in more general
the effective hadronic couplings in the off-shell region, the strong phases
of the effective couplings, and the nonfactorization parameter
$\chi(\mu)$, both of which are needed to be studied in some
nonperturbative methods based on QCD in the future.

\section{Summary} We have studied some channnels of $D\to PV $ decays. The
total decay amplitude includes direct weak decays and final state rescattering
effects. The direct weak decays are calculated in factorization
approach, and the final state interaction effects are studied in
one-particle-exchange method. The prediction of naive factorization
is far from the experimental data. After including the contribution
of final state interaction, the theoretical prediction can accommodate
the experimental data well. The strong phases of the effective hadronic
couplings are neccessary to reproduce experimental data.

\vspace{5mm}

\noindent {\large{\bf Acknowledgement}}\vspace{0.3cm}

\noindent This work is supported in part by National Natural Science
Foundation of China.

\end{document}